\documentclass[pdflatex,sn-nature]{sn-jnl}%

\usepackage{graphicx}%
\usepackage{multirow}%
\usepackage{amsmath,amssymb,amsfonts}%
\usepackage{amsthm}%
\usepackage{mathrsfs}%
\usepackage[title]{appendix}%
\usepackage{xcolor}%
\usepackage{textcomp}%
\usepackage{manyfoot}%
\usepackage{booktabs}%
\usepackage{algorithm}%
\usepackage{algorithmicx}%
\usepackage{algpseudocode}%
\usepackage{listings}%
\usepackage{longtable}
\usepackage{array}
\usepackage[normalem]{ulem}
\usepackage{amssymb}
\usepackage{changes}
\definechangesauthor[color=magenta]{R2}
\definechangesauthor[color=orange]{R3}
\definechangesauthor[color=blue]{AI}
\usepackage{tikz}
\usetikzlibrary{shapes.geometric,arrows.meta,positioning}

\theoremstyle{thmstyleone}%

\theoremstyle{thmstyletwo}%

\theoremstyle{thmstylethree}%

\begin{document}

\title[Article Title]{Warhead Verification with Neutron Beams and an Analog Information Barrier}

\author[1]{\fnm{Nolan} \sur{Kowitt}}
\author[2]{\fnm{Michael} \sur{Moore}}
\author[1]{\fnm{Kevin} \sur{Sanchez}}
\author[2]{\fnm{Mital} \sur{Zalavadia}}
\author*[1]{\fnm{Areg} \sur{Danagoulian}}\email{aregjan@mit.edu}

\affil[1]{\orgdiv{Nuclear Science and Engineering}, \orgname{Massachusetts Institute of Technology}, \orgaddress{\city{Cambridge}, \postcode{02139}, \state{Ma}, \country{USA}}}

\affil[2]{\orgdiv{Pacific Northwest National Laboratory}, \orgaddress{\city{Richland}, \postcode{99354}, \state{Wa}, \country{USA}}}

\abstract{Neutron Resonance Transmission Analysis has been proposed as an isotope sensitive strategy to verify the authenticity of nuclear warheads and warhead components slated for dismantlement in an arms control verification exercise. To ensure acceptance of this technique by treaty partners, the measurement should minimize information learned about the warhead, including both geometric and isotopic features. Past implementations of transmission analysis acquired and compared neutron time-of-flight spectra. We have developed an analog information barrier proof of concept system where only counts in specific resonance energy windows are observed. The system uses discrete analog components, producing a complete data acquisition and analysis circuit. By limiting the design to a small number of discrete, macroscopic components, the apparatus is intended to substantially reduce the burden of authentication and certification relative to digital data-acquisition systems. This analog measurement technique may enable future ambitious arms control treaties that explicitly stipulate the dismantlement of nuclear weapons.}

\keywords{arms control, NRTA, warhead verification}

\maketitle

\section{Introduction}\label{sec1}
The most recent data exchanges between the United States and Russia via the New START treaty  indicated that a combined 2,893 nuclear warheads were deployed by these countries~\cite{state_new_start_treaty,state2022_newstart_aggregate}. New START, which expired in February, 2026, only accounted for the nuclear weapons considered to be strategic, i.e., capable of long range delivery, and those arms that were in active deployment. Non-deployed strategic nuclear warheads, tactical weapons, and those nuclear weapons considered to be retired were neither limited nor counted by the treaty, adding further to the total tally. Proponents of nuclear deterrence argue that the scale of the deployed strategic nuclear forces provides a credible deterrent against direct or major acts of aggression, though the adequacy and stability of this posture remain contested questions of strategic judgment rather than settled fact. However, in the event of nuclear escalation, these weapons will endanger the world population.  A recent study has estimated that there could be more than 5 billion deaths in the event of an all out nuclear war between US and Russia, due to global food insecurity and famine~\cite{xia2022global}.

The need for effective arms control has been episodically acknowledged in official U.S. policy, as in the 2022 U.S. Nuclear Posture Review, which stated

\begin{quote}
    The United States will pursue a comprehensive and balanced approach that places a renewed emphasis on arms control, non-proliferation, and risk reduction to strengthen stability, head off costly arms races, and signal our desire to reduce the salience of nuclear weapons globally. Mutual, verifiable nuclear arms control offers the most effective, durable and responsible path to achieving a key goal: reducing the role of nuclear weapons in U.S. strategy.
\end{quote}
While such priorities shift between administrations, the need for verification technologies that protect sensitive design information is also recognized by sustained multilateral and non-governmental efforts, notably the International Partnership for Nuclear Disarmament Verification -- a partnership of more than 25 states developing a toolkit for disarmament verification -- and the Nuclear Threat Initiative~\cite{ipndv}.
With the expiration of New START, there is also growing concern that, in the absence of new treaties, nuclear arsenals may expand rather than decline. Treaties that could enable verified dismantlement of nuclear weapons would therefore represent a significant advancement in global stability. However, verification technologies capable of confirming the presence of a nuclear weapon, or successful dismantlement, have the simultaneous burden of requiring sufficient confidence in the authenticity of the weapon or process while protecting sensitive design information.

Historically, arms control agreements between the United States and the Soviet Union/Russia emphasized verified dismantlement of delivery vehicles—such as ballistic missiles and strategic bombers—rather than nuclear warheads themselves. This focus was in part motivated by the assumption that delivery systems provide a reliable proxy for a state’s nuclear strike capacity. It also reflects the long standing challenge of negotiating and technically implementing direct verification of nuclear explosives, without revealing sensitive information about their designs~\cite{Hecker}. Safeguarding nuclear weapon design information is paramount to a state's maintenance of an effective deterrent and for preventing unauthorized access. Unintentional release of such sensitive information -- for example, weapon yield-to-mass ratios, miniaturization techniques, or other design parameters -- could shorten the technical path for other state or non-state actors to develop or improve nuclear weapons.  This could undermine a state's confidence in its own deterrent if adversaries gain the ability to model its weapons more precisely.  Such outcomes risk destabilization among nuclear-armed states and broader proliferation~\cite{Hecker}.

One paradigm for warhead verification has been developed by the US national laboratories and is known as template verification. Warhead verification via a template involves comparisons between a candidate Treaty Accountable Item (TAI) and a template, or a "golden copy," which is a previously authenticated object. Confidence in a golden copy's authenticity is achieved from situational context, e.g. by extracting the device from a randomly chosen deployed weapon.  For a detailed discussion on the golden copy, as well as the high-level concept of operations of the template verification process, see Ref.~\cite{engel2019physically}, Section Template Verification. The TAI could be a weapon component, such as a pit. The verification exercise needs to achieve three goals: sensitivity to differentiate a genuine TAI from intentional hoaxes; specificity, represented by the ability to confirm the TAI; privacy, represented by the ability to avoid the collection of sensitive information on TAI. If the candidate TAI is found to match the golden copy, then the candidate is confirmed to be authentic via this templating approach. Research efforts by MIT have shown that epithermal neutron beams can be used in a Neutron Resonance Transmission Analysis (NRTA) method to achieve the previously listed three goals in a physically cryptographic manner~\cite{engel2019physically}. Efforts by researchers from MIT and Princeton, as well as more recent collaborative efforts between Pacific Northwest National Laboratory (PNNL) and MIT, have shown that compact, DT-generator-based NRTA methods can be used to differentiate between various fissile isotopes~\cite{klein2021neutron,mcdonald2024neutron}. These concepts have also drawn international engagement: a recent study at the China Institute of Atomic Energy builds on this physically cryptographic line of work with a deep-learning-based interactive zero-knowledge verification protocol~\cite{ciae2025verification}. However, a concern of adoption within potential future arms control agreements for these, and all other spectroscopic techniques, is that the spectral information acquired during such measurements is very data-rich, making it difficult to achieve a definitive proof that it does not contain any sensitive information.

One approach researched and developed by the national laboratories that addresses the problem of excessive information involves the use of Information Barriers (IBs). The concept is to allow invasive measurements to collect sensitive data, but to safeguard that data within an information protection system. Thus, IBs shift the challenge of verification from the TAI to the IB itself: certification of the proper function of the IB, and authentication that no hidden modifications or undeclared functions exist to exfiltrate or falsify data is necessary. In practice, this remains an extremely difficult task, as the IBs are likely to contain integrated circuits with millions of components in the $\sim$50~nm scale – much smaller than the resolution of X-ray imaging techniques~\cite{bushberg2011essential}. To solve this problem, other researchers have proposed a concept of so-called vintage verification, which uses 1980s computers to mitigate the problems of sub-micron scale and digital complexity~\cite{kutt2019vintage}. For a detailed discussion on various paradigms of verification, including IBs, see Ref.~\cite{Yan02092015}.

In the following study, we describe an approach that abandons the digital realm in its entirety. Instead of using digital electronics, the system consists of a limited number ($\sim$500) of analog, centimeter scale analog electrical components: resistors, capacitors, diodes, and bipolar junction transistors. The system filters the detector signal and allows counts only within well specified time-of-flight (ToF) windows. This  system gates on individual isotopic resonances of choice, blocks the rest, and produces simple TTL logic signals. These are then sent to an analog ripple counter, which displays the count in binary notation using an array of LEDs. All these components can be easily verified, e.g. by destructively assaying them or X-raying to verify that no adulteration is present. This approach is designed to restrict the information available to an inspector to the minimum needed to accept or reject a TAI's authenticity -- on- and off-resonance counts only, never a spectrum -- directly supporting the privacy goal defined above. It will thus increase the trustworthiness of the overall system and be more attractive in a treaty verification regime.  This apparatus is an extension of the NRTA-based verification system described in previous work~\cite{engel2019physically}, where an encrypting mask of unknown composition is additionally used in a template verification approach as a way of making it impossible for inspectors to infer weapon parameters while still being able to compare a golden copy TAI to a candidate TAI. In this study, this mask takes the form of a foil of high-assay low-enriched uranium (HALEU) of undisclosed thickness placed upstream of the TAI. Its role and placement are described in Sec.~\ref{sec:tof_windows}.

DT-generator-based NRTA  isotopic discrimination~\cite{klein2021neutron,mcdonald2024neutron} and physically cryptographic template verification using an encrypting foil~\cite{engel2019physically} have been demonstrated prior to this study. The contribution of the present study is the first physical realization and end-to-end experimental demonstration of a complete measurement and information-restriction chain -- ToF gating, pulse-height discrimination, and counting -- built without any digital logic, firmware, or programmable component anywhere in the signal path. This replaces the digitizer and supporting digital computational framework on which all prior NRTA-based verification concepts, including Ref.~\cite{engel2019physically}, have relied to record spectral information. It removes the associated digital-hardware authentication burden from the entire verification chain.

The experimental setup described in this study is discussed in Section~\ref{sec:methods}, and is similar to the one used in Ref.~\cite{klein2021neutron}. The core methodology behind the measurement system, NRTA, has been thoroughly discussed in the previous literature~\cite{tsuchiya2023development,tang2024nondestructive,zalavadia2026isotopicmeasurementssnmusing,mcdonald2024neutron,klein2021neutron,klein2023neutron,bourke2016non,cj2026pilot,koizumi2024demonstration}. By using the ToF of the neutrons one can perform transmission spectroscopy.  Since many actinides have resonances in the ev-keV epithermal regime, their presence manifests itself as absorption lines in the transmission spectra. In this experiment, we use highly enriched uranium (HEU) plates as a proxy for the so-called golden copy and depleted uranium (DU) plates as a proxy for a hoax TAI. We compare the various measurements of the HEU golden copy, showing clear agreement and thus demonstrating specificity. We also compare the HEU measurements to the DU hoax measurement, showing disagreement and thus demonstrating sensitivity. In the process, we only observe counts from two ToF windows, ensuring privacy.

Figure~\ref{fig:overview} gives a conceptual overview of the full measurement chain described in the remainder of this manuscript, including the role of the digitizer as a calibration and validation aid that is not part of the information-restricted measurement path.

\begin{figure}[h]
    \centering
    \resizebox{\linewidth}{!}{%
    \begin{tikzpicture}[
        node distance=6mm and 7mm,
        box/.style={rectangle, draw, rounded corners, align=center, minimum width=2.2cm, minimum height=9mm, font=\scriptsize},
        arr/.style={-{Latex[length=2mm]}},
        darr/.style={-{Latex[length=2mm]}, dashed}
    ]
        \node[box] (source) {DT Neutron\\Generator};
        \node[box, right=of source] (foil) {HALEU\\Encrypting Foil};
        \node[box, right=of foil] (target) {TAI\\(Target)};
        \node[box, right=of target] (detector) {Detector};
        \node[box, below right=16mm and -33mm of detector, minimum width=3.0cm] (circuit) {Analog Information Barrier\\(ToF gating, SCA, counting)};
        \node[box, right=of circuit] (counts) {Count Display\\(LEDs)};
        \node[box, left=38mm of circuit, minimum width=2.4cm] (digitizer) {Digitizer\\(calibration / normalization /\\ validation only)};
        \draw[arr] (source) -- (foil);
        \draw[arr] (foil) -- (target);
        \draw[arr] (target) -- (detector);
        \draw[arr] (detector.south) -- (circuit.north);
        \draw[arr] (circuit) -- (counts);
        \draw[darr] (digitizer) -- node[above, align=center, font=\tiny] {disconnected\\during exercise} (circuit);
    \end{tikzpicture}%
    }
    \caption{Overview of the measurement chain. Solid arrows show the information-restricted measurement path used during a verification exercise; the dashed path shows the digitizer, used only for pre-measurement calibration, count-rate normalization, and validation (Secs.~\ref{sec2} and~\ref{sec:robustness}), and disconnected during the exercise itself.}
    \label{fig:overview}
\end{figure}

\section{Results}\label{sec:results}
\subsection{Circuit Description}\label{sec2}
\begin{figure}[h]
    \centering
    \includegraphics[width=0.99\linewidth]{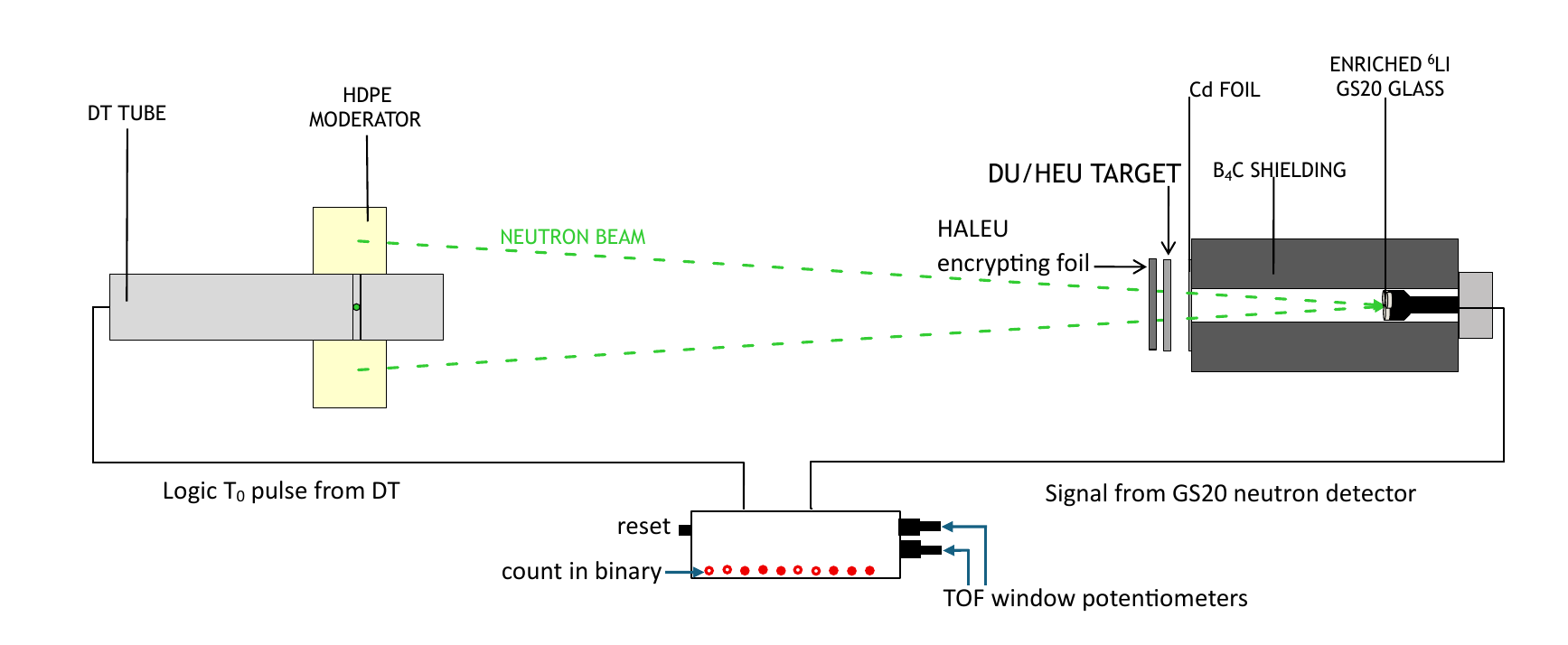}
    \caption{Schematic of the experimental setup. The DT neutron generator uses a radial moderator
composed of high-density polyethylene. The moderated neutron beam is incident on a target and detected in a 5~mm thick GS20 enriched $^6$Li
glass scintillator coupled to a photomultiplier tube. The detector is shielded from off-axis neutrons by boron carbide. Thermal wraparound neutrons are filtered by
2.0 mm of cadmium foil placed along the axis. Modified with permission from Ref.~\cite{klein2021neutron}, Copyrighted to the American Physical Society.}
    \label{fig:setup}
\end{figure}

\begin{figure}[h]
    \centering
    \includegraphics[width=\linewidth]{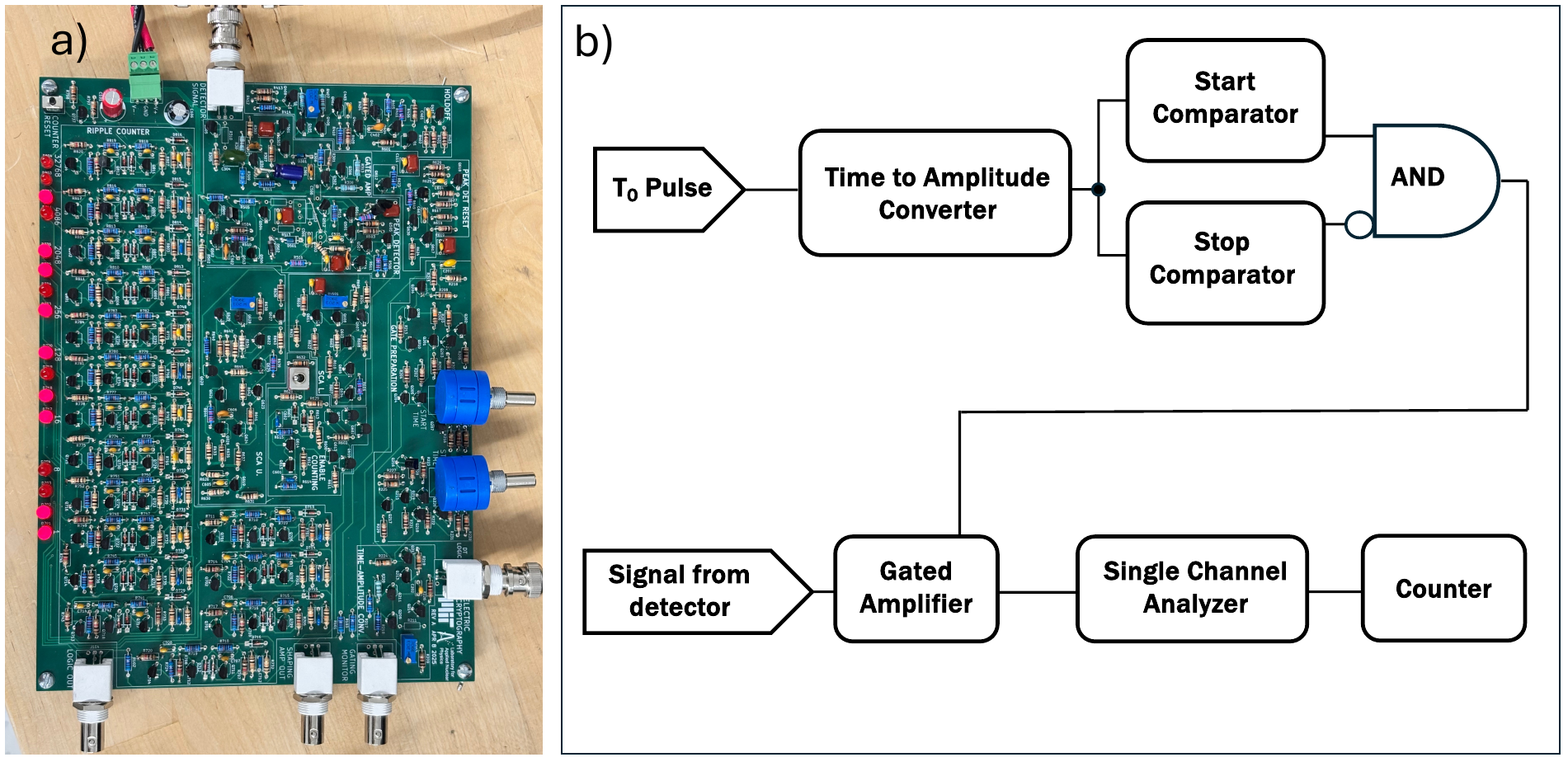}
    \caption{Layout of analog information barrier. a) Photograph of the information barrier circuit in use. The LED row shows the output of the ripple counter, in this case displaying \texttt{0b0010110110110011 = 11699} counts.   b) Schematic diagram showing the functional blocks of the information barrier circuit. See Supplementary Information for a full circuit description. }
    \label{fig:blockdiagram}
\end{figure}

NRTA uses ToF to identify the neutron energies that are preferentially absorbed by a nuclear transition channel. In ToF measurements, windows of time following the neutron generator pulse correspond to energy windows. These TOF windows may contain isotope-specific resonance absorption lines, allowing for sensitivity to the isotopic makeup of the target. The information barrier circuit described in this study is able to measure counts within a ToF window without the use of any digital instrumentation. The physical arrangement of the neutron source, target, and detector is shown in Fig.~\ref{fig:setup}. A schematic representation of the circuit is shown in Fig.~\ref{fig:blockdiagram}. A general description of the circuitry follows below. A detailed description of the circuitry and methods can be found in Supplementary Notes 2-6. Full circuit design of the apparatus, as well as the LTSpice simulation model, are available online in Ref.~\cite{kowitt_github}.

This work employs two distinct mechanisms that together provide the information security of the overall measurement, and we distinguish them explicitly here to avoid conflating them. The analog circuitry described in this section is an information-restriction mechanism: it discards, at the hardware level, all detector signals outside pre-agreed ToF windows before that signal can be recorded, so that counts from only two previously agreed upon ToF bins are ever available to an inspector. It does not by itself hide any property of the TAI. The cryptographic function -- preventing an inspector from inferring the TAI's geometric or isotopic parameters even from the restricted counts that are measured -- is provided separately, by the HALEU encrypting foil of unknown composition described in Ref.~\cite{engel2019physically} and used as an upstream mask in Sec.~\ref{sec:tof_windows}. The two mechanisms are complementary and layered, forming a defense in depth.

\subsubsection{ToF window}
To create a time of flight window during which pulses are measured, a timing signal must be created and synchronized with the neutron generator. This is achieved with a time to amplitude converter (TAC)~\cite{knoll2010radiation}. It consists of a capacitor that is grounded on one end and sourced with a constant current, so that its voltage rises at a constant rate. This capacitor is discharged when the generator produces the neutron pulse, so that its voltage is proportional to the time that has elapsed since the neutron pulse ended. The charging process of this capacitor converts time to voltage. This voltage is buffered and simultaneously measured by two comparators, marked as Start Comparator and Stop Comparator in Fig.~\ref{fig:blockdiagram}. These comparators' thresholds are set by precision potentiometers so that the time at which they change state with respect to the deuterium-tritium (DT) neutron generator  pulse can be reliably set to within 10~ns. These comparators define the beginning and the end of the allowed ToF window and can be combined using a logical AND~\cite{HorowitzHill2015} to produce a single logic pulse corresponding in time to the ToF window. Before the actual experimental measurement, the potentiometers are set by sending the AND signal to an oscilloscope. Then, using the oscilloscope’s built in measurement function, the timing gates can be set by changing the potentiometer values so that the AND pulse matches the desired ToF window. The ToF window(s) should be pre-determined as part of the treaty negotiations. After the ToF window has been set, the oscilloscope and all other digital devices are disconnected from the system. See Supplementary Notes 2 and 3 for more detail.

\subsubsection{Gated Amplifier}
The first step of the signal path is a switch type linear gate~\cite{kowalski1970nuclear}, which amplifies and outputs the detector signal if and only if a logic signal allows it. This is achieved by terminating and buffering the signal and then amplifying it using a common emitter amplifier. The gain of this inverting amplifier is proportional to the ratio of the collector impedance to the emitter impedance~\cite{HorowitzHill2015}.

Outside the ToF window, when the logic pulse is off, a MOSFET that is separated from the collector by a capacitor is brought into saturation. This makes the amplifier's collector resistance at signal frequencies close to zero, meaning that any detector pulses are vetoed by shorting them to ground. The gated amplifier is the only place in the circuits where raw spectral data in the form of analog detector pulses, potentially containing sensitive information, may be present. By carefully verifying these 15 components during the process of authentication and certification, operators can have high confidence that extra information is not learned by inspectors.

During the ToF window, the MOSFET is cut off, meaning that the collector impedance is dominated by the collector resistor, and the gain is nonzero and constant. The detector signal is then passed through and sent to the analysis block of the circuit. %
For a full description of this block of the circuit, see Supplementary Notes 5 and 6.

\subsubsection{Analysis Hardware}
The goal of the subsequent circuitry is to identify pulses consistent with neutron events and count only those pulses. The neutron detector used in this study, GS20, is a $^6$Li doped glass. It is sensitive to both gammas and neutrons. The light emitted from neutron capture is proportional to the $Q$ value of n+$^6$Li$\rightarrow \alpha + ^3$H$+Q$, where $Q=4.785$~MeV. This $Q$ is then converted to scintillation light, resulting in a peak in the pulse height spectrum. Neutron pulses can be separated from the majority of $\gamma$'s by applying a pulse height window; the circuit that does this is known as a single channel analyzer (SCA)~\cite{kowalski1970nuclear}. The threshold values can be adjusted using trim potentiometers, as shown in Fig.~S6, and once set they do not need readjustment throughout an experiment. Since pulse height discrimination is used in multiple detector types, this circuitry can easily be adapted to other detectors and is not limited to GS20. The output of the SCA passes through a switch, which enables and disables counting. The final step of the analysis chain is a series of T flip-flops, each tied to an LED. This binary counter can be reset at run start using a button, and then total counts can be measured and recorded along with the total time of the measurement.  See Supplementary Note 6 for more detail.

\subsubsection{Authentication and Certification Considerations}\label{sec:authentication}

The strength of the system described in this study is based on two important properties.

First, the entire signal path comprises on the order of 500 discrete, centimeter-scale components (resistors, capacitors, diodes, comparators, and bipolar junction transistors), each of which can be individually checked by X-ray imaging, destructive assay or decapsulation, or bench-level electrical testing against a declared reference schematic and bill of materials. This stands in contrast to a digital ASIC, FPGA, or microcontroller-based data-acquisition system, in which confirming the absence of undeclared logic among millions of transistors at the $\sim$50~nm scale is not achievable with current inspection techniques~\cite{Yan02092015}. A circumstance that further strengthens this is the choice of very common and broadly available electronic components, e.g. 2N3904 and 2N3906 BJT transistors and 5\% resistors.

Second, the location within the circuit where raw, potentially sensitive analog detector pulses are present is small and enumerable: it is limited to the 15 components of the gated amplifier described in Sec.~\ref{sec2} above. Every other stage of the circuit -- the ToF window logic, the single channel analyzer, and the ripple counter -- carries only binary gating signals or accumulated counts, neither of which reveal spectral shape. A digital DAQ system, by comparison, typically retains sensitive pulse data in memory, storage, and firmware states that are not amenable to the same kind of direct physical inspection.

The two properties of this design reduce the authentication burden to a specific, finite set of macroscopic components, rather than to an integrated circuit whose full state space would be difficult to verify with the current state of the art technologies. This claim has limits that the present study does not resolve. Discrete analog circuits are not inherently immune to tampering or hardware Trojans; a malicious or substituted component could in principle be introduced at fabrication or during shipping.  This study does not define a certification protocol -- including provenance requirements on the supplied hardware, the level of destructive testing required, or the frequency of re-certification -- nor does it address authentication of the broader measurement chain beyond the analog processing board itself, including the detector, photomultiplier tube voltage dividers, high-voltage supply, cabling, and neutron generator interface.  These problems should be left for future studies.  The results of this study show that the authentication burden for this apparatus is significantly smaller and more tractable than that of a comparable digital system. Equipment and component provenance, and confidence measures for those components is an additional open question. We expect this question to be addressed the same way as the ToF windows themselves (Sec.~\ref{sec2}): through treaty-level negotiation between the parties, analogous to existing provisions for establishing a golden copy in the template verification paradigm~\cite{engel2019physically}.

\subsection{Experimental Demonstration}
The analog information barrier circuit has been constructed and used in several measurements. The digital spectra shown in this study were only used for validation and benchmarking of the analog circuit. All conclusions of this study are solely based on standalone measurements from the analog apparatus. In a real verification exercise, inspectors can select any combination of these windows to measure, contingent on prior agreements.

\subsubsection{Verification of HEU using ToF windows}\label{sec:tof_windows}
\begin{figure}[h!]
    \centering
    \includegraphics[width=\linewidth]{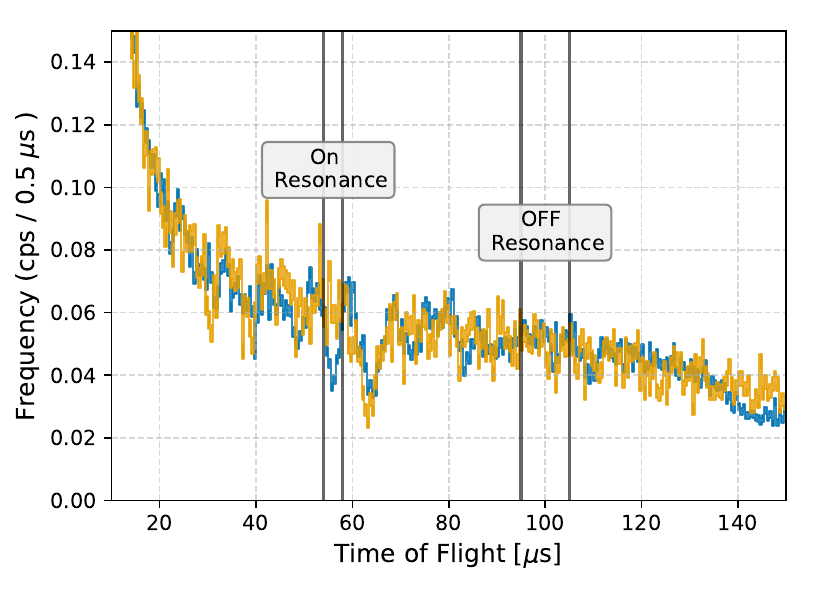}
    \caption{Time of flight spectrum for Uranium targets. Spectra for DU and HEU targets were acquired with a CAEN DT5790 digitizer. The analog cryptographic system only produces counts in the marked energy windows.}
    \label{fig:TOF_HEU_DU}
\end{figure}

The analog information barrier was demonstrated at Pacific Northwest National Laboratory (PNNL), and used to differentiate a hoax proxy device from a golden copy device. To emulate a verification exercise, two targets were assembled and separately tested. The first target consisted of a 0.7~mm thick 91.7\% $^{235}$U-enriched HEU foil encased in 2 mm of polycarbonate~\cite{zalavadia2026isotopicmeasurementssnmusing}. This served as a proxy for a golden copy and an honest TAI candidate. In addition to this, a sheet of 0.5~mm thick 19.8\%-enriched high assay low enriched uranium (HALEU) strips attached to an aluminum substrate were placed upstream of the HEU, to emulate the effect of an encrypting mask, as a way of replicating the physically cryptographic technique described in Ref.~\cite{engel2019physically}. The HEU assembly was then replaced with an assembly containing a 2 mm thick depleted uranium (DU) plate. This DU assembly emulated a hoax scenario. The DU target also contained a polyethylene puck designed to match the hydrogen content of the polycarbonate HEU case. Digitized epithermal neutron transmission spectra, acquired with a CAEN DT5790 digitizer DAQ system are shown in Fig.~\ref{fig:TOF_HEU_DU}.

The spectra show qualitatively similar behavior because the broadband neutron count rate is determined by the amount of low-Z polycarbonate in the target. Both exhibit a deep resonance around 63~\textmu s, which corresponds to the 6.7 eV resonance in $^{238}$U~\cite{Zhang2019iNEUIT}. In order to differentiate the targets, it was decided to measure the count rate within the 8.8 eV resonance in $^{235}$U, which corresponds to 56~\textmu s.

The analog information barrier measures the total counts within a ToF window marked up in Fig.~\ref{fig:TOF_HEU_DU}. The resonance window was chosen to be centered on the 8.8~eV resonance of \textsuperscript{235}U and was set from 54~\textmu s to 58~\textmu s using the two comparator threshold potentiometers.  To prevent a hoaxing scenario where $^{235}$U is replaced  with a broadly absorbing material, a second part of the spectrum was also measured between 95~\textmu s and 105~\textmu s. This second window was chosen because neither isotope had a resonance there. By dividing the on-resonance count rates by the off-resonance count rates, a normalized resonant depth could be determined. It monotonically decreases as the thickness of $^{235}$U increases.

The results of the test are presented in Table~\ref{table:final data}. Combining the results of HEU measurements across two days, we determine ratio of $r_\text{HEU}=0.3720 \pm 0.0081$. We then compare this to the $r_\text{DU}=0.482 \pm 0.016$ using a $Z$-test: $Z=(r_\text{DU}-r_\text{HEU})/\sqrt{0.0081^2+0.016^2}=6$. In this exercise, we can then reject the identity between the DU and HEU samples with the confidence of 6$\sigma$, which corresponds to a $p$-value of $p=10^{-9}$. At the same time the HEU measurements show statistically identical results of $r_\text{HEU}$, thus proving the specificity of the technique.
As can be seen, the system is stable across days. When the amount of $^{235}$U in the target was lowered via the DU hoax, the system was able to identify the change. Table~\ref{table:final data} is a comprehensive summary of all observed results and represents the totality of information available to an inspector. By applying predetermined statistical tests to this ratio, it can be used to verify or reject the identity between a golden copy and a TAI candidate with high confidence. This demonstrates the feasibility of using this analog cryptographic apparatus for performing verification. While this exercise compared the two extremes of enrichment, the measured on/off-resonance ratio decreases monotonically with the areal density of $^{235}$U in the beam, so a TAI of intermediate enrichment would produce an intermediate ratio, distinguishable from both the template and the hoax given sufficient counting statistics.

To further verify that the analog information barrier is working correctly, the digitized spectra were also analyzed, and counts and ratios were determined. The ratios, as determined from the digitized spectra, were found to be 0.362~±~0.009 and 0.489~±~0.02 for HEU and DU, respectively. These numbers match the values determined using the analog results shown in the last column of Table~\ref{table:final data}, showing that the apparatus is working correctly.

\subsubsection{Verification by Mapping of a Resonance}
Rather than measuring counts inside on- and off-resonance windows as described above, inspectors could alternatively verify the isotopic composition of a TAI based on multiple measurements spanning a single resonance. As the thickness of the TAI increases, due to the high peak cross sections of the resonances, the bins of the spectrum corresponding to the center of the resonance lose areal density sensitivity.  However, it has been shown in the study in Rev.~\cite{engel2019physically} that even in these scenarios NRTA can be sensitive to changes in the composition of thick targets, due to the partial attenuation in the off-center shoulders of the resonance, where the cross section is lower. Compared to measuring transmission only across wide TOF windows, as demonstrated in Section~\ref{sec:tof_windows}, mapping a single resonance may be advantageous; because each isotope's resonance broadens uniquely with increasing areal density, the detailed resonance shape provides robustness against both isotopic and geometric hoaxing. The primary disadvantage is the increased measurement time: mapping a resonance with multiple narrow windows requires acquisition times roughly proportional to the number of windows used.

\begin{figure}[!h]
    \centering
    \includegraphics[width=0.8\linewidth]{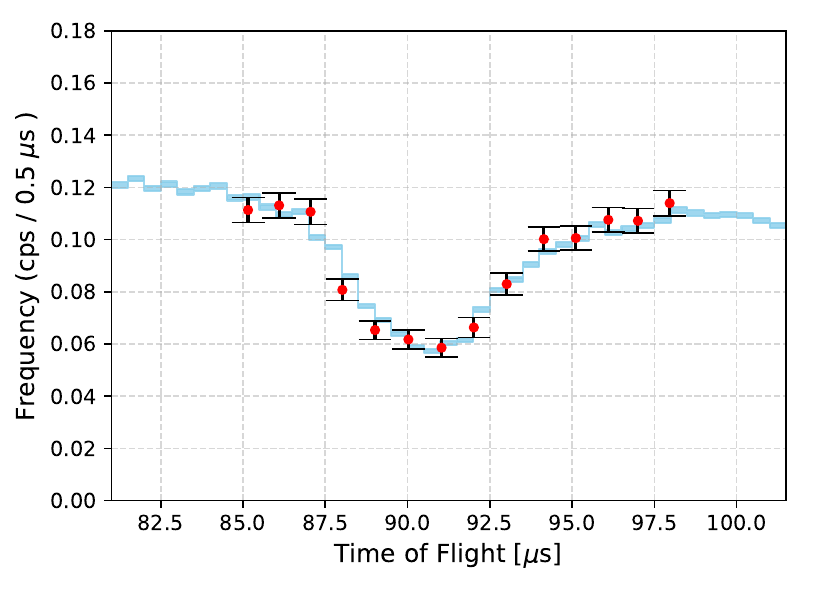}
    \caption{Time of flight measurement of Tungsten target. TOF measurements of the 4.3 eV resonance of $^{182}$W, an 8-hour digitized spectrum is presented as the ground truth, for comparison with the separate 40-minute analog measurements. For the analog measurements, the horizontal width represents the timing window measured, while the magnitude of the errorbars are ±1 standard deviation. The results show good agreement ($\chi^2/\text{NDF} = 9.08/14$).}
    \label{fig:TOFresolved}
\end{figure}

To confirm that the circuit possesses sufficient spectroscopic resolution to support this approach and resolve the shapes of the resonances, a series of measurements were performed in which the counting window was shifted across a resonance between consecutive runs. A 1~mm thick W plate was interrogated at MIT using neutrons produced by a Thermo Fisher P383 DT generator. The ToF spectrum was first recorded with a benchtop digitizer to establish a ground truth for testing the analog circuit. The same target was then independently measured using the analog information barrier, both measurements shown in Fig.~\ref{fig:TOFresolved}. The 4.3 eV resonance of $^{182}$W was chosen as a test case because tungsten is accessible in an academic setting and its resonance, with a peak cross section of 10~kbarn, is fully black at this thickness. For comparison, most \textsuperscript{235}U and \textsuperscript{239}Pu resonances in the 1-20~eV regime are no more than 2~kbarn. The resulting analog count rates, recorded in narrow windows across the resonance, are shown overlaid on the digitized spectrum in Fig. 3. The circuit resolves the resonance and agrees with the digitized spectrum within measurement uncertainties ($\chi^2/\text{NDF} = 9.08/14$). This demonstrates that the circuit has the spectroscopic resolution necessary to make the resonance-mapping measurement feasible.

\subsection{Inspection Protocol}\label{sec:protocol}

The inspection protocol implemented by this apparatus follows the template verification exercise described in detail in Ref.~\cite{engel2019physically}, specialized to on/off-resonance counting rather than full-spectrum comparison. In outline: (1) the genuine reference object (the ``golden copy''), together with an encrypting foil, is measured first to establish the template on/off-resonance ratio; (2) the candidate TAI is measured under identical, jointly witnessed conditions, using the same ToF windows and encrypting-foil arrangement; (3) the two measurements are compared using a pre-agreed statistical test
whose outcome determines whether the candidate is accepted as genuine or rejected as a hoax. Figure~\ref{fig:protocol} summarizes this exercise. This directly specifies both how confidence is established from the inspector's perspective and how acceptance criteria are set: the criterion is the statistical test itself, with its significance threshold, the ToF windows, and the measurement time fixed in advance by treaty negotiation, as described earlier in Sec.~\ref{sec2}.

Spoofing resistance is addressed directly by the results already presented: Sec.~\ref{sec:tof_windows} demonstrates rejection of an isotopic hoax (DU substituted for HEU, 6$\sigma$) and acceptance of honest HEU candidates. Geometric-hoax resistance, which Ref.~\cite{engel2019physically} addresses through measurements under multiple random rotations, has not yet been tested with this analog apparatus and remains for future work.

Equipment trust is addressed in Sec.~\ref{sec:authentication} above; here we note a concrete, precedented mechanism from Ref.~\cite{engel2019physically} directly applicable to that discussion: the inspecting party can be asked to supply $n$ copies of the instrumentation, of which the host destructively validates $n-2$, leaving the remaining two to be used interchangeably during the exercise. This bounds the probability of an undetected malicious substitution at $p=2/[n(n-1)]$,
which can be made arbitrarily small by increasing $n$.

\begin{figure}[h]
    \centering
    \begin{tikzpicture}[
        node distance=8mm,
        box/.style={rectangle, draw, rounded corners, align=center, minimum width=6.2cm, minimum height=9mm, font=\small},
        decision/.style={diamond, draw, aspect=2.5, align=center, font=\small, inner sep=1pt},
        arr/.style={-{Latex[length=2mm]}}
    ]
        \node[box] (negotiate) {Treaty negotiation: ToF windows,\\ measurement time, statistical threshold};
        \node[box, below=of negotiate] (template) {Measure genuine reference object\\ (golden copy) with agreed encrypting foil\\ $\rightarrow$ template on/off-resonance ratio};
        \node[box, below=of template] (candidate) {Measure candidate TAI\\ under identical, jointly witnessed conditions};
        \node[box, below=of candidate] (test) {Pre-agreed statistical test\\ (e.g. Z-test on resonance ratio)};
        \node[decision, below=of test] (decision) {Passes\\ threshold?};
        \node[box, below left=10mm and -8mm of decision] (accept) {Candidate accepted\\ as genuine};
        \node[box, below right=10mm and -8mm of decision] (reject) {Candidate rejected\\ as hoax};
        \draw[arr] (negotiate) -- (template);
        \draw[arr] (template) -- (candidate);
        \draw[arr] (candidate) -- (test);
        \draw[arr] (test) -- (decision);
        \draw[arr] (decision) -| node[near start, above, font=\scriptsize] {yes} (accept);
        \draw[arr] (decision) -| node[near start, above, font=\scriptsize] {no} (reject);
    \end{tikzpicture}
    \caption{Schematic of the overall proposed inspection protocol. This protocol is described in Sec.~\ref{sec:protocol}, adapted from the template verification exercise described in Ref.~\cite{engel2019physically}.}
    \label{fig:protocol}
\end{figure}

\subsection{Operational Robustness}\label{sec:robustness}

The DT generator exhibited day-to-day output instability: raw on- and off-resonance count rates varied by more than 15\% across identical configurations on a single day (Supplementary Note: Variable Neutron Production Rate). The 6$\sigma$ discrimination result reported in Table~\ref{table:final data} uses only the two days (Sep.~2 and 4) that did not exhibit this instability; it does not depend on the digitizer-based normalization used to correct the excluded, more unstable dataset. %
Day-to-day reproducibility is supported by agreement between the two included HEU measurements ($r=0.372\pm0.024$ vs.\ $0.374\pm0.01$), and the comparator thresholds defining the ToF windows are set to within 10~ns using an oscilloscope before each measurement. As a proof of concept, this study does not characterize potentiometer drift, detector gain drift, or the frequency at which timing and pulse-height thresholds would need re-calibration over longer deployments; these efforts belong to the later design stages of the final inspection instrument.

\section{Discussion}\label{sec3}
We have demonstrated that the method of analog information barrier can securely and reliably measure portions of a ToF window and use them to reject a DU hoax with high confidence when comparing it to an HEU item in a template verification paradigm. Furthermore, the technique clears honest HEU items. In a verification scenario, parties must agree ahead of time on the windows to be measured, measurement times, and statistical tests to be used. By combining the results of several measurements, as demonstrated above, the information from the separate ToF windows could be used to reliably confirm the authenticity of a TAI. This can be achieved without revealing any other isotopic information about the warhead.

Furthermore, similar to prior studies~\cite{engel2019physically}, this study used encrypting foils of HALEU to protect isotopic information about the TAIs.  Combining the new analog technique with the physically cryptographic method creates a more robust technique and a defense in depth for information security. While the measurements presented here used uranium, the method is not specific to it: $^{239}$Pu and $^{240}$Pu possess strong isotope-specific resonances in the same few-eV epithermal region, with peak cross sections below that of the tungsten resonance used to demonstrate the spectroscopic resolution of the circuit in Sec.~\ref{sec:results}. The apparatus and protocol therefore carry over to plutonium-bearing TAIs unchanged, with only the ToF windows repositioned to the relevant resonances; an experimental demonstration with plutonium, which was not available for this campaign, remains for future work.

It is useful to place this approach against other proposed strategies for restricting the release of sensitive information during warhead verification. Purely digital DAQ/MCA-based NRTA systems, including our own prior work~\cite{klein2021neutron,mcdonald2024neutron}, record the full time-of-flight spectrum digitally, which affords greater post-hoc analysis flexibility but requires certifying that all digital signal-processing and storage hardware in the chain faithfully applies the agreed information restriction and contains no additional, undeclared logic. Information Barrier concepts built on modern digital integrated circuits~\cite{Yan02092015} face a similar burden: verification of sub-micron-scale digital hardware is not presently amenable to complete physical inspection (Sec.~\ref{sec:authentication}). "Vintage verification" approaches, which use well-characterized 1980s-era digital computers to reduce this problem~\cite{kutt2019vintage}, ease but do not eliminate it, since these systems still rely on programmable logic and stored software state that must itself be authenticated. The fully analog circuit described here has no software state and no digital field-programmable elements at any stage of the measurement or analysis chain. Furthermore, all digital platforms include components in the sub-\textmu m scale that are currently impossible to test individually.  The corresponding trade-off is flexibility: both modern-digital and vintage-digital approaches can be reconfigured to measure different ToF windows or resonances without rebuilding hardware, whereas our approach requires physical reconfiguration of circuit components to do so.

Future iterations of the analog information barrier method could use a single TAC, with several time gated amplifiers all operating in parallel. This would allow inspectors to simultaneously measure count rates within multiple ToF windows while maintaining the information security of the overall method. This would reduce the time required for a verification measurement at the cost of increased circuit complexity. Additional improvements, based on feedback from subject matter experts, would involve general improvements in the circuitry,  the use of larger components, as well as limiting the PCB to no more than two layers. These would simplify the process of authentication and certification~\cite{whyatt_private_comm}.

It should also be noted that the test objects used here are geometrically simple compared to a realistic TAI, which would involve more complex special nuclear material geometries. These would produce more complex transmission spectra and could require re-selecting the off-resonance window used to normalize the resonance signal, while variations in counting geometry between template and candidate measurements could elevate false-negative rates. Characterizing these effects with more representative test objects is an important direction for future work and should be part of the design stages of the final instrument.  These considerations concern the verification protocol rather than the apparatus itself, since the analog circuit is agnostic to target geometry and its ToF windows are modifiable.

A verification system based on analog components alone would vastly improve the ability to authenticate its use. Such a device would essentially amount to the bottom row of the diagram presented in Fig.~\ref{fig:blockdiagram}(b). Simplifying the certification and authentication burden for inspection equipment ahead of their use with nuclear weapons significantly reduces the security challenges of implementing new technology. To that end, the approach presented in this study can be adapted within other verification strategies. For example, the SCA of the circuit described here could be modified for use with gamma spectroscopy measurements, such as those proposed for dismantlement transparency ~\cite{moore2024applied}, or neutron ratio and multiplicity counting techniques ~\cite{marleau2023joint}. However, no state has yet attempted to verifiably confirm the dismantlement of nuclear weapons within an arms control agreement with another party. In order to first demonstrate, then negotiate, and finally deploy new confirmation technologies to support global stability, novel practical solutions to restrict the potential release of sensitive design information must be adopted.

\section{Methods }\label{sec:methods}

\subsection{Experimental Setup}

The basic DT-generator-based NRTA technique has been previously described in Ref.~\cite{klein2021neutron}. In this particular study, we used the setup that is described in great detail in Ref.~\cite{mcdonald2024neutron,zalavadia2026isotopicmeasurementssnmusing}. The schematic of the experimental setup is shown in Fig.~\ref{fig:setup}. The distance between the downstream face of the moderator and the upstream face of the detector was measured with a Leica DISTO E7400x to be 2.001m. The ThermoFisher P385 was operated with an accelerating voltage of 130~kV and a beam current of 35 \textmu A. The generator was pulsed with a repetition rate of 5 kHz and a 3.5\% duty factor. While this translates to a nominal pulse width of 7~\textmu s, due to the limitations of the DT generator the pulse width was estimated to be approximately 1~\textmu s~\cite{zalavadia2026isotopicmeasurementssnmusing,mcdonald2024neutron}.  A gaussian fit to the pulse resulted in a $\sigma = 0.55$~\textmu s. This narrow pulse was crucial in ensuring the ToF spectroscopic resolution necessary for this study.

\subsection{Neutron Detector}
The neutron detector used in this study consisted of a PMT-mounted GS20 scintillator. GS20 is a $^6$Li doped glass produced by Scintacor Ltd.~\cite{scintacor_li6_glass}. It is sensitive to both gammas and neutrons. The light emitted from neutron capture is proportional to the $Q$ value of n+$^6$Li$\rightarrow \alpha + ^3$H$+Q$, where $Q=4.785$~MeV. Due to quenching effects, this energy corresponds to approximately 1.6~MeV\textsubscript{ee}~\cite{oshima2011temperature}.  This results in a peak in the pulse height spectrum, allowing it to be isolated from the continuum generated by gammas by placing a narrow integration cut.  In the analog circuit, this is achieved by defining the two thresholds in the SCA block via two trim potentiometers.  Detailed testing of the analog circuit was performed, where the potentiometer settings were changed to count within different values of the pulse height spectrum.  This was then compared to the digitized spectra, showing strong agreement. For details of this procedure see Supplementary Information.

\section*{Data Availability}
Source data are provided with this paper. The data is available online in the repository at Ref.~\cite{kowitt_github}.

\section*{Code Availability}
The LTSpice circuit simulation model and full circuit design files used in this study are available online in Ref.~\cite{kowitt_github}. Analysis code is available from the corresponding author upon request.

\bibliography{sn-bibliography}%

\section*{Acknowledgments}
The Claude Sonnet 5  large language model was used for word editing and clarity;  all edits then underwent thorough checks by the authors.

\section*{Funding Statement}
This work was in part funded by the NNSA NA-221 award DE-NA0003920. Nolan Kowitt was supported by a Founder's Fellowship from the Department of Nuclear Science and Engineering, MIT.

\section*{Author Contributions Statement}
NK: methodology, investigation, writing, MM: investigation and Writing – review and editing, KS: investigation, MZ: investigation, AD: Conceptualization, writing, supervision. All authors reviewed and approved the final manuscript.

\section*{Competing Interests Statement}
The authors declare no competing interests.

\section*{Tables}
\begin{table*}[ht]
\caption{Results of the HEU and DU measurements from the analog information barrier. The Ratio column shows a clear, statistically significant discrimination between HEU and DU targets. This table represents the totality of information available to an inspector in a verification exercise. It shows a clear difference between DU and HEU samples.}
\resizebox{\textwidth}{!}{
\begin{tabular}{ccccccccc}
\hline
\textbf{Date} & \textbf{Target} & \textbf{\shortstack{On-resonance \\ counts /time [s]}} & \textbf{On rate [cps]} & \textbf{\shortstack{Off-resonance \\ counts /time [s]}} & \textbf{Off rate} & \textbf{Ratio} \\
\hline\hline
2 Sep & HEU & 0x146 / 720.4 & 0.45 ± 0.03 & 0x3fe / 840.25 & 1.22 ± 0.04 & 0.372 ± 0.024 \\
2 Sep & HEU & 0x2c3 /1570.6 & 0.45 ± 0.02 & 0x49e / 960.3 & 1.23 ± 0.04 & 0.366 ± 0.017 \\
4 Sep & HEU & 0x911 / 4424 & 0.53 ± 0.01 & 0xe81 / 2645 & 1.40 ± 0.02 & 0.374 ± 0.01 \\
4 Sep & DU & 0x567 / 2207 & 0.63 ± 0.02 & 0xb66 / 2243 & 1.30 ± 0.02 & 0.482 ± 0.016 \\
\hline
\end{tabular}
\label{table:final data}
}
\end{table*}

\section*{Figure Legends/Captions}

\begin{enumerate}
\item Overview of the measurement chain. Solid arrows show the information-restricted measurement path used during a verification exercise; the dashed path shows the digitizer, used only for pre-measurement calibration, count-rate normalization, and validation (Secs.~\ref{sec2} and~\ref{sec:robustness}), and disconnected during the exercise itself.

\item Schematic of the experimental setup. The DT neutron generator uses a radial moderator
composed of high-density polyethylene. The moderated neutron beam is incident on a target and detected in a 5~mm thick GS20 enriched $^6$Li
glass scintillator coupled to a photomultiplier tube. The detector is shielded from off-axis neutrons by boron carbide. Thermal wraparound neutrons are filtered by
2.0 mm of cadmium foil placed along the axis. Modified with permission from Ref.~\cite{klein2021neutron}, Copyrighted to the American Physical Society.

\item Layout of analog information barrier. a) Photograph of the information barrier circuit in use. The LED row shows the output of the ripple counter, in this case displaying \texttt{0b0010110110110011 = 11699} counts.   b) Schematic diagram showing the functional blocks of the information barrier circuit. See Supplementary Information for a full circuit description.

\item Time of flight spectrum for Uranium targets. Spectra for DU and HEU targets were acquired with a CAEN DT5790 digitizer. The analog cryptographic system only produces counts in the marked energy windows.

Legend: Blue = Digitized HEU Spectrum
Yellow: Digitized DU Spectrum

\item Time of flight measurement of Tungsten target. TOF measurements of the 4.3 eV resonance of $^{182}$W, an 8-hour digitized spectrum is presented as the ground truth, for comparison with the separate 40-minute analog measurements. For the analog measurements, the horizontal width represents the timing window measured, while the magnitude of the errorbars are ±1 standard deviation. The results show good agreement ($\chi^2/\text{NDF} = 9.08/14$).

Legend: Blue = Ground truth, digital
Red points = Analog

\item Schematic of the overall proposed inspection protocol. This protocol is described in Sec.~\ref{sec:protocol}, adapted from the template verification exercise described in Ref.~\cite{engel2019physically}.

\end{enumerate}

\end{document}


\title[Article Title]{Supplementary Information: Warhead Verification with Neutron Beams and an Analog Information Barrier}

\author[1]{\fnm{Nolan} \sur{Kowitt}}
\author[2]{\fnm{Michael} \sur{Moore}}
\author[1]{\fnm{Kevin} \sur{Sanchez}}
\author[2]{\fnm{Mital} \sur{Zalavadia}}
\author[1]{\fnm{Areg} \sur{Danagoulian}}

\affil[1]{\orgdiv{Nuclear Science and Engineering}, \orgname{Massachusetts Institute of Technology}, \orgaddress{\city{Cambridge}, \postcode{02139}, \state{Ma}, \country{USA}}}

\affil[2]{\orgdiv{Pacific Northwest National Laboratory}, \orgaddress{\city{Richland}, \postcode{99354}, \state{Wa}, \country{USA}}}

\maketitle

\section*{Supplementary Note 0. Validation Measurements}\label{Full_Spectrum}
To verify the operation of the system, a Thermo Fisher P383 neutron generator was used to make several measurements of a 0.5~mm tungsten plate on the MIT campus. For these measurements, the DT generator was operated at 110~kV accelerating voltage, with a beam current of 40~\textmu A. The generator was pulsed at 5~kHz with a 3.2\% duty factor, and placed 2.04~m from the detector. Both a CAEN DT5720 digitizer and the analog information barrier were used to measure the TOF spectrum in separate acquisitions, with no changes to the experimental configuration between them.

\begin{figure}[h!]
    \centering
    \includegraphics[width=\linewidth]{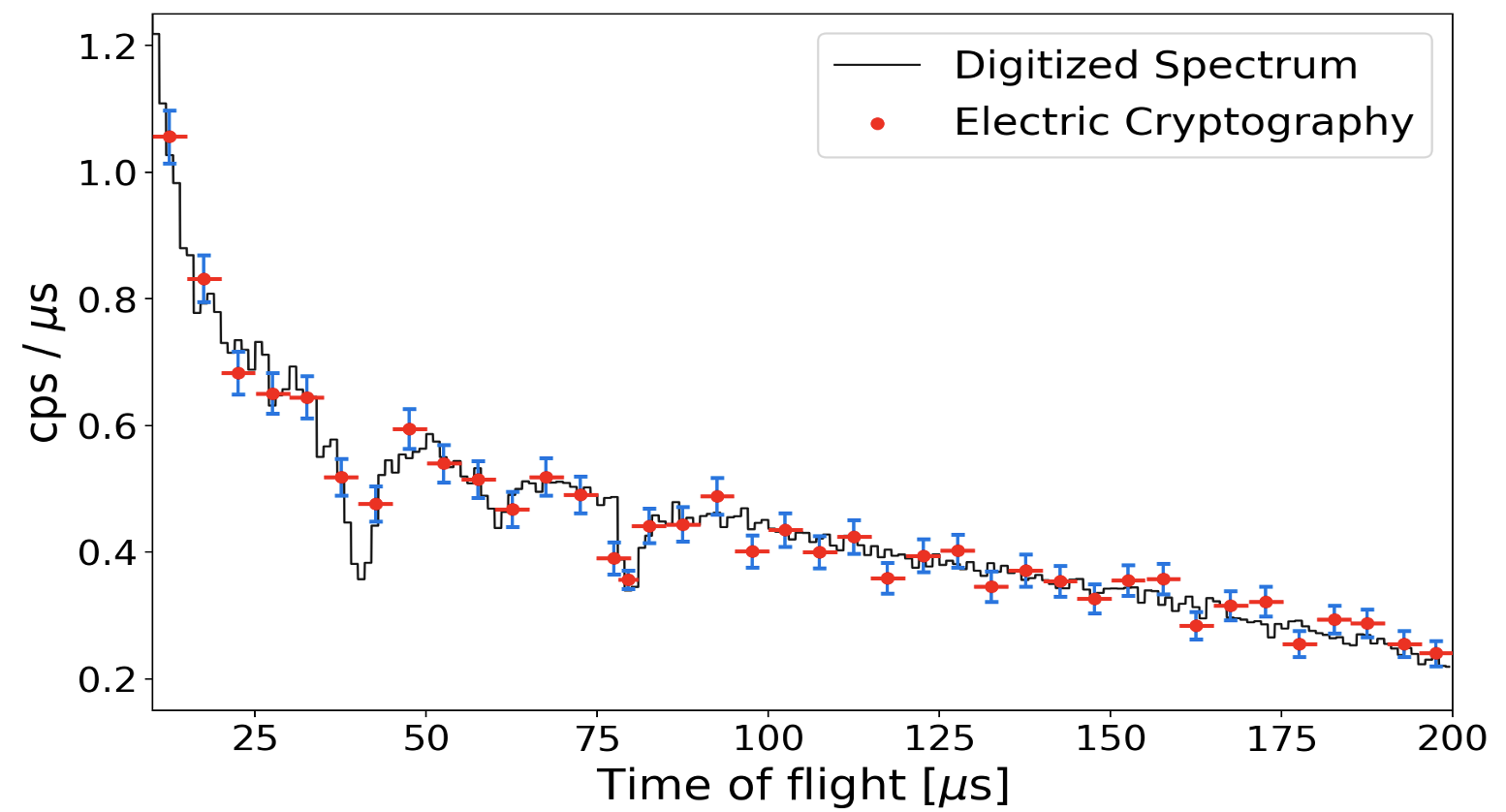}
    \caption{TOF spectrum of a W target, measured for one hour with a benchtop digitizer and in separate 2-minute measurements with the analog information barrier. For the analog measurements, the vertical error bars represent the count rate uncertainty, while the horizontal bars show the counting window for each measurement.}
    \label{fig:TOFW}
\end{figure}

Each point in Fig.~\ref{fig:TOFW} is the average neutron count rate within the TOF window, normalized by the width of the window for comparison with the digitized spectrum. The TOF windows were set to have a width of 5 \textmu s, with the actual extent of the counting window indicated by the horizontal red bars. The uncertainty is dominated by counting statistics, so better agreement can be achieved with longer counting periods. A comparison of the digitized spectrum with the analog measurements gave a reduced $\chi^2$ value of 0.926, indicating good agreement between the two spectra~\cite{hughes2010measurements} .

\section*{Supplementary Note 1. Variable Neutron Production Rate}

\begin{table*}
\resizebox{\textwidth}{!}{
\begin{tabular}{ccccccccc}
\hline
\textbf{Date} & \textbf{Target} & \textbf{\shortstack{On resonance \\ counts /time [s]}} & \textbf{On rate [cps]} & \textbf{\shortstack{Off resonance \\ counts /time [s]}} & \textbf{Off rate} & \textbf{Ratio} \\
\hline\hline
2 Sep & HEU & 0x146 / 720.4 & 0.45 ± 0.03 & 0x3fe / 840.25 & 1.22 ± 0.04 & 0.372 ± 0.024 \\
2 Sep & HEU & 0x2c3 /1570.6 & 0.45 ± 0.02 & 0x49e / 960.3 & 1.23 ± 0.04 & 0.366 ± 0.017 \\
3 Sep & HEU & 0x496 / 2574.3 & 0.46 ± 0.01 & 0x7b8 / 1682.2 & 1.18 ± 0.03 & 0.388 ± 0.014 \\
3 Sep & DU & 0x356 / 1449 & 0.59 ± 0.02 & 0x7ad / 1511 & 1.30 ± 0.03 & 0.453 ± 0.019 \\
3 Sep & HEU & 0x4fe / 2338 & 0.55 ± 0.02 & 0xce4 / 2356 & 1.40 ± 0.02 & 0.390 ± 0.013 \\
4 Sep & HEU & 0x911 / 4424 & 0.53 ± 0.01 & 0xe81 / 2645 & 1.40 ± 0.02 & 0.374 ± 0.01 \\
4 Sep & DU & 0x567 / 2207 & 0.63 ± 0.02 & 0xb66 / 2243 & 1.30 ± 0.02 & 0.482 ± 0.016 \\
\hline
\end{tabular}
}
\caption{Summary of measurements acquired at PNNL. The effects of generator output instabilities are most clearly seen when comparing the raw on-resonance and off-resonance count rates, which varied by over 15\% for identical configurations on the same day.}
\end{table*}

During the measurement campaign at PNNL, the DT generator output exhibited some time-varying instabilities. On September 3, instabilities in the DT generator output increased significantly, resulting in poor agreement with the expected ratios. Once the issue was identified, it was addressed by simultaneously acquiring data with both the information barrier and a digitizer. While the purpose of the analog circuit is to avoid digital systems, the digitizer was only used to measure the total number of detected pulses for normalization purposes, and no spectroscopic data was used in the analysis. In future work, this count rate normalization will be implemented in analog hardware. By dividing both measured count rates by the total detection rate recorded by the digitizer, the effects of this drift were eliminated. After applying this correction, the DU and HEU ratios were 0.465 ± 0.019 and 0.382 ± 0.013, respectively. Although this approach temporarily required digital instrumentation for normalization, an analog implementation using a parallel counting channel can achieve the same result. By measuring on-resonance and off-resonance simultaneously, the measurement would be immune to the effects of a slowly changing neutron source.

\section*{Supplementary Note 2. TAC}\label{TAC_Description}
For the circuit description here and in the rest of the supplement, component labels refer to the actual component names on the PCB~\cite{kowitt_github}.

The goal of the time-to-amplitude converter (TAC) is to produce a voltage proportional to the time elapsed since the end of the DT pulse. When the DT generator is pulsing, it outputs a 5~V logic signal with 50~\ohm output impedance. This signal drives Q214 into saturation, discharging C202. As soon as the DT pulse ceases, Q214 is cut off and can be neglected for the remainder of the analysis.

Transistors Q209 and Q211 form a cascoded current source. The base of Q209 is biased with a stiff voltage divider, so that it is always at about 0.9 times the value of V+, which in this case is 12~V. The voltage across the emitter resistance, R211$\parallel$(R209+RV202), is approximately $V_+\cdot(1-0.9)-V_{BE}$, resulting in a constant collector current. The combination of fixed resistors and trim potentiometer allows the effective emitter resistance, and thus the charging current, to be adjusted. The collector current of Q209 is passed through Q211 to charge C202. As C202 charges, Q209's $V_{CE}$ will remain constant, shielding Q209 from both the Early effect and most temperature fluctuations~\cite{HorowitzHill2015}. Following the end of the DT pulse, the voltage across the capacitor will be: $$V(t) = \int_0^t{\frac{I}{C}}dt = t\cdot\frac{V_+\cdot(1-0.9)-V_{BE}}{R_E\cdot C202}+V_0$$

where $V_0$ is the saturation voltage of Q214 and $R_E$ is the total emitter resistance used to set the current. The linear rise in voltage means that the TAC amplitude sensitivity, $\frac{dV}{dt}$, is uniform across the entire TOF window. The downside of this approach, compared to a simple RC-based TAC, is that the current source saturates once the capacitor voltage reaches $V_+ - 2\cdot V_{BE}$, at which point the voltage ceases to rise.The adjustable emitter resistance, however, allows the charging current to be tuned so that the linear region of the TAC spans the full period of the neutron generator.

To use the TAC output, the voltage must be buffered. This is achieved using Q212 and Q213, which are configured as emitter followers, so that Q213 appears to the integrating capacitor as a $\beta\cdot 10 k\Omega\approx 1$M\ohm load. There are two followers, one made with an NPN and the other made with a PNP, so that their constant $V_{BE}$ offsets will approximately cancel each other. This complementary transistor configuration is employed in several places throughout the circuit to minimize offset voltages.

\begin{figure}[h]
    \centering
    \includegraphics[width=\linewidth]{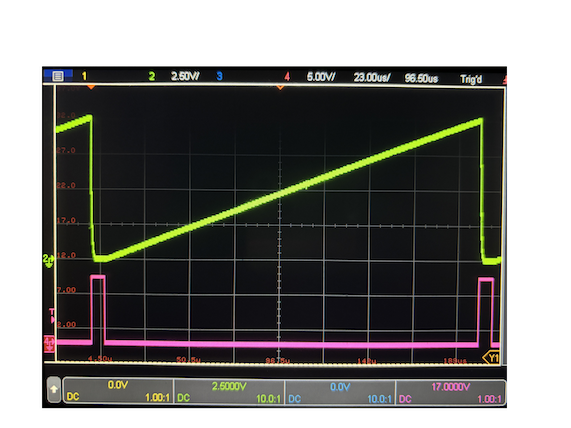}
    \caption{Osilloscope trace showing the TTL pulse from the DT generator (pink) as well as the buffered TAC output (yellow).}
    \label{fig:TAC_Scope_Trace}
\end{figure}

\section*{Supplementary Note 3. Comparators}\label{comp_Description}
Once the TAC has been buffered, comparators are required to establish a TOF window. The basic design of the comparators is the long-tailed pair~\cite{HorowitzHill2015}, formed by transistors Q204 and Q205 (and duplicated with Q218 and Q219 for the second threshold). The comparator design is duplicated for the start and end of the window, with the inverting and noninverting inputs swapped. Operated without feedback, the high gain causes the output to swing sharply when the noninverting input exceeds the inverting input. The reference voltage, which establishes the TOF window, is set using a 10-turn trim potentiometer. This potentiometer, RV201, forms a voltage divider that drives the base of Q205. When the TAC voltage is lower than Q205's base, Q204 is cut off, and Q205's collector is near ground. Once Q204's base rises above that of Q205, Q204 conducts, raising the shared emitter voltage and cutting off Q205. The tail current is set by Q207, configured as a current sink. This active biasing raises the differential gain, sharpening the switching transition and providing common-mode rejection so that the comparator threshold does not shift at different TAC voltages. The output of the differential amplifier is further amplified by two common emitter amplifiers. Finally, the output of the comparator is buffered by an emitter follower (Q202) to drive subsequent stages.
\newline

\section*{Supplementary Note 4. Gated Amplifier}\label{Gated_amp_Description}
One of the most important aspects of the circuit is the gated amplifier, as it is the only stage where detector pulses outside the TOF window are present.

The input stage of the amplifier is an emitter follower, used to buffer the detector anode signal. The detector signal can  be terminated either externally using a BNC tee or by populating resistor R310.

The core of the gated amplifier is a common emitter amplifier built with Q304. The base of Q304 is biased by the resistor divider R304/R309, with the buffered anode signal AC-coupled to it through C304. The coupling time constant must be chosen so that $C304\cdot(R304\parallel R309)\gg \tau_{pulse}$ to avoid pulse distortion. At signal frequencies (when the gate is open), the emitter resistance is dominated by R312, and the gain of the amplifier is~$-(\frac{R303\parallel R307}{R311\parallel R312})$. Similarly, C302 must satisfy $C302\cdot R307\gg \tau_{pulse}$. The compensation capacitor C303 rolls off the gain of the amplifier at high frequencies, preventing oscillations.

When the logic pulse is high, the channel conduction of Q301 is high, reducing the collector load at signal frequencies. This reduces the gain of the amplifier to near zero, effectively suppressing detector pulses that arrive outside the TOF window and excluding them from subsequent analysis stages. Q303 is an emitter follower that buffers the gated amplifier output.

\begin{figure}
    \centering
    \includegraphics[width=0.6\linewidth]{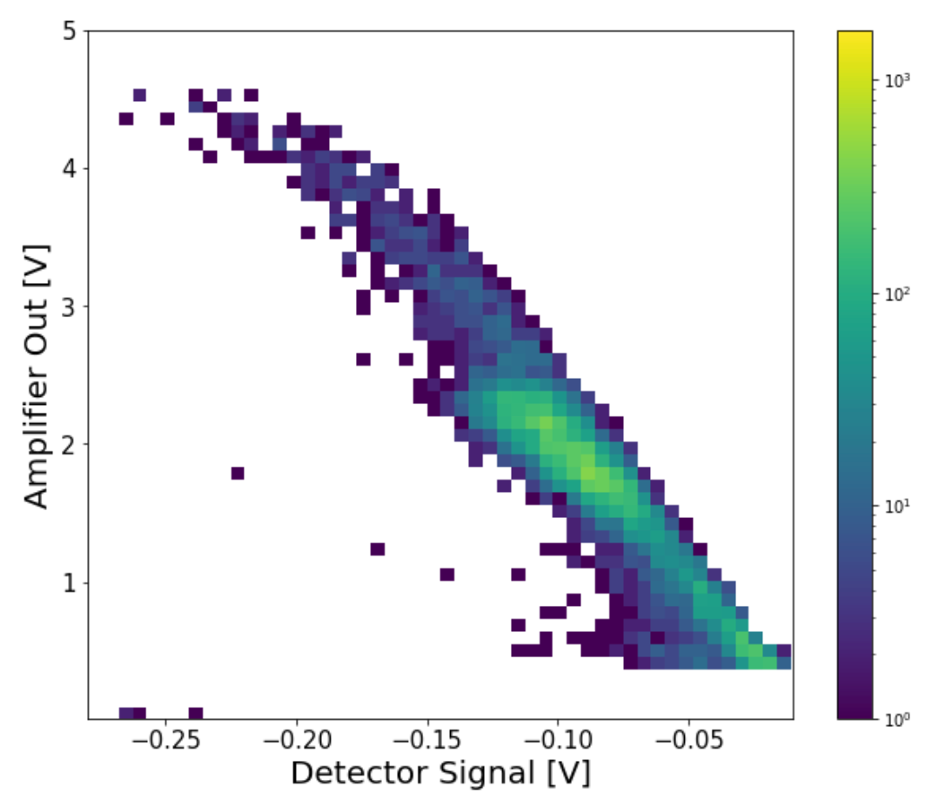}
    \caption{Measured gain of the gated amplifier when the gate is open, where the colorbar represents the total number of counts measured. }
    \label{fig:ampgain}
\end{figure}

Electronic switching noise of approximately 200~mV was observed on the output of this amplifier but did significantly impact performance given the pulse height of the neutron capture peak. In the measurements of neutrons shown in Fig. S6, the neutrons produced pulse heights between 1.6 and 2.6~V.

\section*{Supplementary Note 5 Single Channel Analyzer}\label{SCA_Description}
The Single Channel Analyzer (SCA), which was designed specifically for this project, comprises several functional blocks. At the heart of the SCA is a peak detector, whose output follows the rising edge of pulses, but maintains the voltage after the input falls. The output of this peak detector is read by two comparators, a lower voltage comparator which corresponds to the minimum pulse heights (LLD), and an upper voltage comparator that sets the maximum pulse height (ULD). To prevent prematurely triggering, a holdoff circuit is activated at the start of the rising edge. This holdoff holds the SCA output low until the pulse has finished rising, ensuring that the peak value is captured before the comparators evaluate the pulse.

\subsection*{5.1: Pulse Preparation}
When a new pulse arrives, the first thing that must happen is the generation of a holdoff pulse, to prevent the SCA from firing on pulses whose amplitudes exceed the ULD. The front end is a high-gain differential amplifier in which one input (the base of Q406) is held at a fixed reference voltage. This value can be set once using fixed resistor values, and so long as it lies above the switching noise and below the minimum neutron detection pulse height, will reliably trigger on neutrons. This logic pulse is buffered by an emitter follower to avoid loading the differential amplifier. The buffered signal is then AC-coupled into a monostable pulse generator~\cite{HorowitzHill2015}. The pulse generator produces a holdoff pulse of approximately 3.5~\textmu s duration, set by C402 and R405. This duration should be minimized while remaining longer than the longest expected pulse rise time, as the holdoff dominates the system dead time.

In the signal path following the gated amplifier is a small pulse stretcher, made with a diode and capacitor. The discharge time constant of the stretcher is relatively short compared to the holdoff duration. By slightly lengthening the pulse duration to about 3~\textmu s, the stretcher provides the peak detector sufficient time to slew to the true peak value.
\subsection*{5.2: Peak Detector}

The peak detector design, a precision diode-capacitor circuit with an active feedback loop, was adopted from [3] (p. 225). To avoid using operational amplifiers, however, the feedback amplifiers were implemented as discrete high-gain differential stages. The Darlington pairs, composed of pairs Q510 and Q515, are effectively single transistors with a current gain equal to $\beta^2$. The load on the voltage storing capacitor C501 is thus equal to the impedance of the current source multiplied by $\beta^2$. This high input impedance ($\beta^2 \cdot R_{current\space source}$) was necessary to prevent droop of the stored voltage during the holdoff period. There is a nonzero bias current that slowly charges C501, which would eventually register as a false peak. This is mitigated by R513, which provides a discharge path for the bias current.

\begin{figure}[h]
    \centering
    \includegraphics[width=0.6\linewidth]{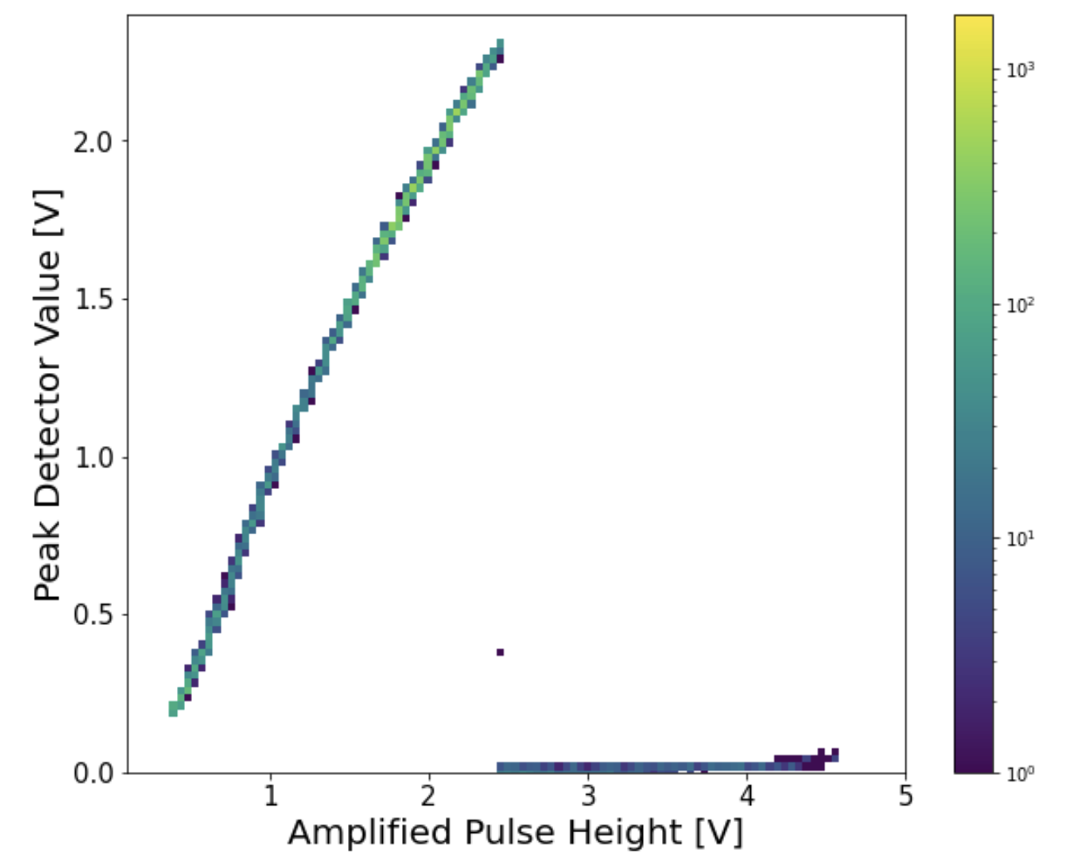}
    \caption{Measured output voltage of the peak detector at the time when the holdoff pulse ends, i.e., the moment at which the SCA decision is made. Note the SCA upper threshold was set to 2.6~V, which causes the abrupt drop in peak detector outputs.}
    \label{fig:pkdet}
\end{figure}

The two MOSFETs Q612 and Q620 are nominally non-conducting but discharge C501 whenever either gate is driven high. Q612 is used to discharge the capacitor following a pulse detected at the output of the SCA. This resets the peak detector so that it is low when the next pulse arrives. Q620 is controlled by the upper-level comparator via a pulse generator similar to that in the holdoff circuit. This holds the peak detector low as soon as a pulse passes above the ULD, for a duration exceeding the pulse width, ensuring the peak detector remains reset throughout. This prevents the output of the SCA from going high, and means that by the time the holdoff expires, the peak detector will be low again and the SCA will be ready to accept another pulse, without ever producing a count.

Two comparators identical in design to those described in Section 3 implement the ULD and LLD SCA thresholds. They were evaluated by keeping the upper threshold 100~mV above the lower, and scanning this window across the dynamic range of the system. In doing this a differential pulse height spectrum is measured, which qualitatively agrees with measurements acquired independently with a CAEN digitizer.

\begin{figure}[h!]
    \centering
    \includegraphics[width=0.6\linewidth]{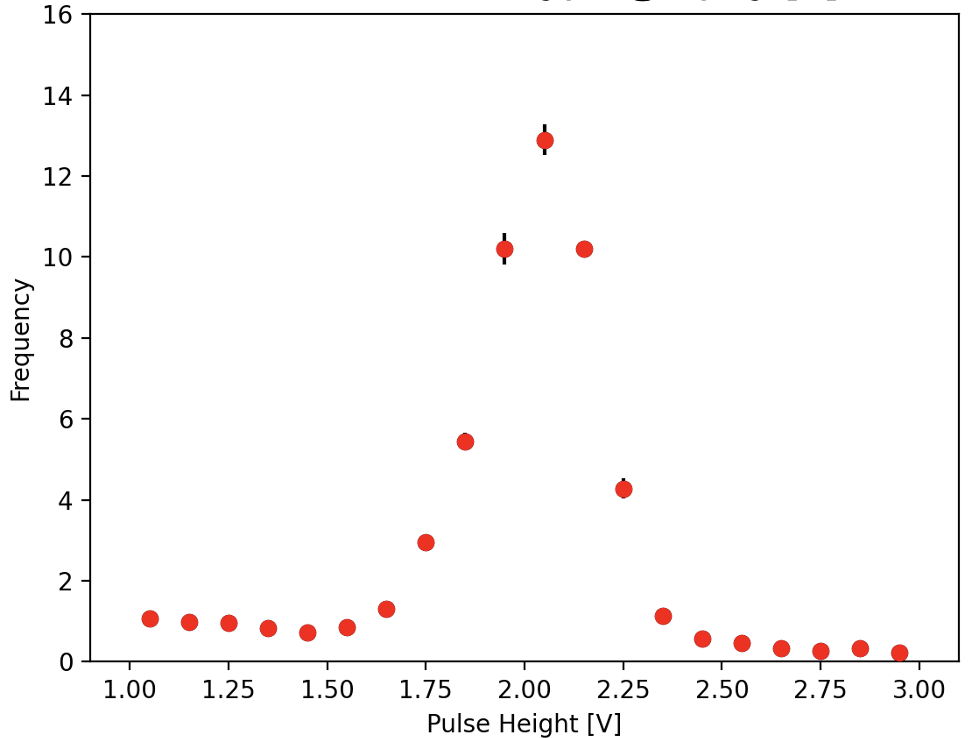}
    \caption{Differential pulse height spectrum measured with a DT neutron source, showing the single neutron capture peak.}
    \label{fig:neutronpeak}
\end{figure}

\subsection*{5.3: Counter}\label{Counter_Description}
The final stage of the signal chain is a counter, so that the SCA pulses can be accumulated over a measurement time. To define the measurement interval precisely, the SCA output is gated with a debounced toggle switch through a logical AND. Pulses are counted only while the switch is in the 'on' state. The operator records the elapsed counting time with a stopwatch, enabling a count rate measurement. A momentary push-button with a pull-up resistor provides a manual reset, initializing all flip-flops to zero before each measurement.

The operation of the flip-flop was adapted from~\cite{KABtroniks2011}. Each flip-flop controls a transistor which is either cut off or in saturation. When in saturation, collector current flows through an LED, illuminating it to indicate a logic '1' state. Although the present implementation was designed and manufactured with 16 bits, an arbitrary number can be strung in series.

\section*{Supplementary References}
\begingroup
\renewcommand{\refname}{}
\bibliography{sn-bibliography}
\endgroup